\documentclass[a4paper]{article}

\usepackage{amsmath}

\begin{document}

\title{\textbf{Singularity analysis of a new discrete nonlinear Schr\"{o}dinger equation}}

\author{\textsc{Sergei Sakovich}\medskip\\{\small Institute of Physics, National Academy of Sciences of Belarus,}\\{\small 220072 Minsk, Belarus. E-mail: saks@tut.by}}
\date{}
\maketitle

\begin{abstract}
We apply the Painlev\'{e} test for integrability to a new discrete (differ\-ential-difference) nonlinear Schr\"{o}dinger equation introduced by Leon and Manna. Since the singular expansions of solutions of this equation turn out to contain nondominant logarithmic terms, we conclude that the studied equation is nonintegrable. This result supports the observation of Levi and Yamilov that the Leon--Manna equation does not admit high-order generalized symmetries. As a byproduct of the singularity analysis carried out, we obtain a new discrete equation which should be integrable according to a conjecture of Weiss.
\end{abstract}

\section{Introduction}

Leon and Manna \cite{LM} introduced the following new discrete (differential-differ\-ence) nonlinear Schr\"{o}dinger equation:
\begin{equation}
\alpha \partial_t^2 \psi_{m} = \mathrm{i} \beta \left( \psi_{m+1}-\psi_{m-1} \right) + 2 \left| \psi_{m} \right|^{2} \psi_{m} , \label{psi}
\end{equation}
where $\alpha$ and $\beta$ are nonzero real parameters, $m = 0 , \pm 1 , \pm 2 , \dotsc$. This equation~\eqref{psi} was derived in \cite{LM} from the integrable Toda lattice equation using the reductive perturbation analysis. Later Levi and Yamilov \cite{LY} studied the integrability of \eqref{psi} by means of the generalized symmetry analysis, and found that the equation \eqref{psi} does not admit local generalized symmetries of order higher than three and, hence, does not possess the same integrability properties as of the Toda lattice equation, from which it has been derived.

In the present research note, we study the integrability of the Leon--Manna equation \eqref{psi} by means of the Painlev\'{e} test, following the Ablowitz--Ramani--Segur algorithm of singularity analysis \cite{ARS} (see also the review \cite{RGB}). We discover the presence of nondominant logarithmic terms in the singular expansions of solutions of~\eqref{psi}. This brings us to the same conclusion as made in \cite{LY}, namely, that the new discrete nonlinear Schr\"{o}dinger equation \eqref{psi} must be nonintegrable. As a byproduct of the singularity analysis carried out, we obtain a new discrete equation which should be integrable according to the conjecture formulated by Weiss in \cite{W}.

\section{Singularity analysis}

Following \cite{LY}, we study the equation \eqref{psi} in its equivalent form:
\begin{equation} \label{uv}
\begin{split}
\partial_t^2 u_{m} & - u_{m}^{2} v_{m} - u_{m+1} + u_{m-1} = 0 , \\
\partial_t^2 v_{m} & - v_{m}^{2} u_{m} + v_{m+1} - v_{m-1} = 0 ,
\end{split}
\end{equation}
$m = 0 , \pm 1 , \pm 2 , \dotsc$, where $u_{m}$ and $v_{m}$ stand for $\psi_{m}$ and $\bar{\psi}_{m}$, and a complex-valued scale transformation of variables has been made.

From the standpoint of singularity analysis, the equation \eqref{uv} is an infinite system of second-order nonlinear ordinary
differential equations, which admits infinitely many branches of the dominant behavior of solutions, i.e.\ the choices of constants $\sigma_{m} , \tau
_{m} , u_{m,0} , v_{m,0}$ in $u_{m} = u_{m,0} \phi^{\sigma_{m}} + \dotsb$ and $v_{m} = v_{m,0} \phi^{\tau_{m}} + \dotsb$, where $m = 0 , \pm 1 , \pm 2 , \dotsc$ and $\partial_t \phi = 1$. We will analyze only one of those branches, namely,
\begin{equation}
\sigma_{m} = \tau_{m} = -1 , \qquad u_{m,0} v_{m,0} = 2 , \qquad m = 0 , \pm 1 , \pm 2 , \dotsc . \label{br}
\end{equation}
Using the expansions
\begin{equation} \label{re}
\begin{split}
u_{m} & = u_{m,0} \phi^{-1} + \dotsb + u_{m,r_{m}} \phi^{r_{m}-1} + \dotsb , \\
v_{m} & = v_{m,0} \phi^{-1} + \dotsb + v_{m,r_{m}} \phi^{r_{m}-1} + \dotsb
\end{split}
\end{equation}
with $u_{m,0} v_{m,0} = 2$, we find from \eqref{uv} that the positions of resonances are $r_{m} = -1 , 0 , 3 , 4$, for all $m$, $m = 0 , \pm 1 , \pm 2 , \dotsc$. Though this branch \eqref{br} represents only a special type of the singular behavior of solutions, its analysis is sufficient for reaching the conclusion that the system \eqref{uv} does not pass the Painlev\'{e} test for integrability.

Assuming that the solutions of \eqref{uv} are represented in the case of \eqref{br} by the Laurent type expansions
\begin{equation}
u_{m} = \sum_{i=0}^{\infty} u_{m,i} \phi^{i-1} , \qquad v_{m} = \sum_{i=0}^{\infty} v_{m,i} \phi^{i-1} , \label{ex}
\end{equation}
with $\partial_t \phi = 1$, we find the following recursion relations for the constant coefficients $u_{m,n}$ and $v_{m,n}$, $m = 0 , \pm 1 , \pm 2 , \dotsc$, $n = 0 , 1 , 2 , \dotsc$:
\begin{equation} \label{rr}
\begin{split}
(n-1) (n-2) u_{m,n} & - \sum_{i=0}^{n} \sum_{j=0}^{n-i} u_{m,i} u_{m,j} v_{m,n-i-j} \\
& + u_{m-1,n-2} - u_{m+1,n-2} = 0 , \\[8pt]
(n-1) (n-2) v_{m,n} & - \sum_{i=0}^{n} \sum_{j=0}^{n-i} v_{m,i} v_{m,j} u_{m,n-i-j} \\
& - v_{m-1,n-2} + v_{m+1,n-2} = 0 , \\
\end{split}
\end{equation}
where $u_{m,-2} = v_{m,-2} = u_{m,-1} = v_{m,-1} = 0$ formally.

Now we have to check whether the recursion relations \eqref{rr} are consistent at all the resonances. At $n=0$, we have $u_{m,0} = 2 / v_{m,0}$ with arbitrary nonzero constants $v_{m,0}$, $m = 0 , \pm 1 , \pm 2 , \dotsc$, and no compatibility conditions arise. Next we obtain
\begin{equation}
u_{m,1} = v_{m,1} = 0 \label{n1}
\end{equation}
and
\begin{equation} \label{n2}
\begin{split}
u_{m,2} & = \frac{2}{3 v_{m-1,0}} - \frac{2}{3 v_{m+1,0}} + \frac{v_{m-1,0} - v_{m+1,0}}{3v_{m,0}^{2}} , \\
v_{m,2} & = \frac{v_{m,0}^{2}}{6 v_{m+1,0}} - \frac{v_{m,0}^{2}}{6 v_{m-1,0}} + \frac{v_{m+1,0} - v_{m-1,0}}{3}
\end{split}
\end{equation}
from \eqref{rr} with $n=1$ and $n=2$, respectively; $m = 0 , \pm 1 , \pm 2 , \dotsc$. At the resonance $n=3$, we obtain
\begin{equation}
v_{m,3} = - \frac{1}{2} v_{m,0}^{2} u_{m,3} \label{n3} ,
\end{equation}
where $u_{m,3}$ are arbitrary constants, $m = 0 , \pm 1 , \pm 2 , \dotsc$, and again no compatibility conditions arise. At the highest resonance, $n=4$, however, the recursion relations \eqref{rr} fail to be consistent, and the infinite set of compatibility conditions
\begin{gather}
\frac{v_{m-1,0}^{2}}{v_{m,0} v_{m-2,0}} + \frac{2 v_{m,0}}{v_{m-2,0}} + \frac{2 v_{m-2,0}}{v_{m,0}} + \frac{v_{m,0} v_{m-2,0}}{v_{m-1,0}^{2}} \notag \\
+ \frac{v_{m+1,0}^{2}}{v_{m,0} v_{m+2,0}} + \frac{2 v_{m,0}}{v_{m+2,0}} + \frac{2 v_{m+2,0}}{v_{m,0}} + \frac{v_{m,0} v_{m+2,0}}{v_{m+1,0}^{2}} \notag \\
- \frac{2 v_{m,0}^{2}}{v_{m-1,0} v_{m+1,0}} - \frac{4 v_{m-1,0}}{v_{m+1,0}} - \frac{4 v_{m+1,0}}{v_{m-1,0}} - \frac{2 v_{m-1,0} v_{m+1,0}}{v_{m,0}^{2}} = 0 \label{cc}
\end{gather}
arises there for the infinite set of constants $v_{m,0}$, $m = 0 , \pm 1 , \pm 2 , \dotsc$.

Since we have obtained the nontrivial compatibility conditions \eqref{cc}, we have to modify the singular expansions \eqref{ex} by introducing appropriate logarithmic terms, starting from the terms proportional to $\phi^{3} \log\phi$ \cite{ARS,RGB}. The appearance of such logarithmic terms, however, is generally believed to be a clear symptom of nonintegrability \cite{RGB}. For this reason, we conclude that the differential-difference equation \eqref{psi} should not be expected to be integrable.

\section{Discussion}

We have studied the integrability of the differential-difference equation \eqref{psi} by means of the Painlev\'{e} test. The singularity analysis has brought us in a simple and straightforward way to the same conclusion as made by Levi and Yamilov who used the higher symmetry approach \cite{LY}, namely, that the new discrete nonlinear Schr\"{o}dinger equation \eqref{psi} introduced by Leon and Manna \cite{LM} must be nonintegrable.

There is one more outcome of our research. As a byproduct of the singularity analysis carried out, we have obtained the new discrete (difference) equation \eqref{cc}, which can be written in the following more compact form using the notation $z_{m} = v_{m,0}$, $m = 0 , \pm 1 , \pm 2 , \dotsc$:
\begin{gather}
\frac{z_{m-1}^{2}}{z_{m} z_{m-2}} + \frac{2 z_{m}}{z_{m-2}} + \frac{2 z_{m-2}}{z_{m}} + \frac{z_{m} z_{m-2}}{z_{m-1}^{2}} \notag \\
+ \frac{z_{m+1}^{2}}{z_{m} z_{m+2}} + \frac{2 z_{m}}{z_{m+2}} + \frac{2 z_{m+2}}{z_{m}} + \frac{z_{m} z_{m+2}}{z_{m+1}^{2}} \notag \\
- \frac{2 z_{m}^{2}}{z_{m-1} z_{m+1}} - \frac{4 z_{m-1}}{z_{m+1}} - \frac{4 z_{m+1}}{z_{m-1}} - \frac{2 z_{m-1} z_{m+1}}{z_{m}^{2}} = 0 . \label{de}
\end{gather}
An interesting conjecture formulated by Weiss \cite{W} states that the differential constraints which arise in the singularity analysis of nonintegrable equations are themselves always integrable (see \cite{S,KS} for further discussion on this conjecture). Therefore we predict that the discrete equation \eqref{de} is integrable, and we would like to attract attention of experts in discrete systems to the problem of integrability study for this new equation.

\end{document}